\documentclass[tightenlines,eqsecnum,floats,aps,amsmath,amssymb,nofootinbib,prd,showpacs]{revtex4}
\usepackage{amsmath,amssymb,amsfonts}
\usepackage{graphicx}
\usepackage{enumerate} 
\usepackage{colordvi} 
\usepackage{bm}
\usepackage{subfigure}
\usepackage{epsf}
\newcommand{\simgt}{\lower.5ex\hbox{$\; \buildrel > \over \sim \;$}}
\newcommand{\simlt}{\lower.5ex\hbox{$\; \buildrel < \over \sim \;$}}

\def\texp{{t_{\rm exp}}} 
\begin{document}
\title{Optimizing future imaging 
survey
of galaxies to confront dark energy and modified gravity models}

\author{Kazuhiro Yamamoto}
\affiliation{Graduate School of Science, 
Hiroshima University, 
Higashi-Hiroshima, 735-8526, Japan}
\author{David Parkinson}
\affiliation{Astronomy Centre, University of Sussex, 
Brighton BN1 9QH, United Kingdom}
\author{Takashi Hamana}
\affiliation{National Astronomical Observatory of Japan
2-21-1 Osawa, Mitaka, Tokyo, 181-8588, JAPAN}
\author{Robert C. Nichol}
\affiliation{ICG, University of Portsmouth, Portsmouth, PO1~2EG, 
United Kingdom}
\author{Yasushi Suto}
\affiliation{Department of Physics and Research Center for the Early
Universe, The University of Tokyo, Tokyo 113-0033, Japan}
\begin{abstract}
We consider the extent to which future imaging surveys of galaxies can 
distinguish between dark energy and modified gravity models for the 
origin of the cosmic acceleration.  Dynamical dark energy models may 
have similar expansion rates as models of modified gravity, yet predict 
different growth of structure histories. We parameterize the cosmic 
expansion by the two parameters, $w_0$ and $w_a$, and the linear growth 
rate of density fluctuations by Linder's $\gamma$, independently. Dark 
energy models generically predict $\gamma \approx 0.55$, while the DGP 
model $\gamma \approx 0.68$. To determine if future imaging surveys can 
constrain $\gamma$ within 20 percent (or $\Delta\gamma<0.1$), we perform 
the Fisher matrix analysis for a weak lensing survey such as the 
on-going Hyper Suprime-Cam (HSC) project.  Under the condition that the 
total observation time is fixed, we compute the Figure of Merit (FoM) as 
a function of the exposure time $\texp$. We find that the tomography 
technique effectively improves the FoM, which has a broad peak around 
$\texp\simeq {\rm several}\sim 10$ minutes; a shallow and wide survey is 
preferred to constrain the $\gamma$ parameter.  While $\Delta\gamma < 
0.1$ cannot be achieved by the HSC weak-lensing survey
alone, one can improve the constraints by combining with a follow-up 
spectroscopic survey like WFMOS and/or future CMB observations.
\end{abstract}


\maketitle

\section{Introduction}


The existence of the mysterious cosmic acceleration is usually ascribed to
the presence of an extra component of the universe with a negative
pressure, known as dark energy. However, modification of the law of gravity remains as another
interesting and equally valid possibility. One of the most elaborated examples is the DGP
cosmological model that incorporates the self-acceleration mechanism
\cite{DGP,GG} without dark energy.
A fundamental question in this context is whether it is possible to
distinguish between the modified gravity and dark energy models that
have an (almost) identical cosmic expansion history \cite{YBRSY,Heavens}. 
The answer to the
question is inevitably dependent on the specific model of dark energy or 
modified gravity\cite{Kunz}. Thus we focus on the DGP model, and consider if it
has any observational signature that can be distinguished from dark
energy models with future galactic surveys. While it is pointed out
that the DGP model has some theoretical inconsistency at a fundamental
level\cite{Luty,Nicolis,ghostkoyama}, it is still useful as an empirical
prototype of modified gravity models, and its observational consequences 
are discussed \cite{MM,SSH,KM,YBRSY}.

The important key is the growth rate of cosmological density
perturbations, which should be different in the two models even 
if they have an identical cosmic expansion history. 
The weak lensing power spectrum can be sensitive to the growth rate,
while the uncertainty of the clustering bias will be the bottleneck
that makes the galaxy power spectrum insensitive to the growth rate.

Currently several imaging and spectroscopic surveys of galaxies are
planned to unveil the origin of cosmic acceleration via weak lensing 
and baryon acoustic oscillation methods. The Hyper Suprime-Cam (HSC) 
project is a fully-funded imaging survey at the Subaru telescope, 
which is expected to commission in 2011. An associated spectroscopic 
survey possibility, Wide-field Fiber-fed Multi-Object Spectrograph 
(WFMOS) project, is under serious discussion between Subaru and Gemini
observatories (see e.g. \cite{DETF,RESA} and references therein for
other projects).

In the present paper, we consider the extent to which future imaging and
spectroscopic surveys of galaxies can distinguish between the DGP and
dark energy models. More specifically, we empirically characterize the
growth rate of density fluctuations adopting Linder's $\gamma$
parameter. By optimizing imaging surveys and the combination with
redshift survey following the previous
literature\cite{OptimalLensingTom,David}, we consider how we can
constrain the value of $\gamma$ from HSC weak lensing survey and/or
WFMOS baryon acoustic oscillation (BAO) survey.

The present paper is organized as follows: In section 2, we explain our
theoretical modeling: the parameterization of the background expansion
and the modified gravity, the Fisher matrix analysis of the weak lensing
power spectrum, and the modeling of the galaxy sample. A demonstration
with the DGP model and dark energy model is also presented.  In section
3, our result of the Fisher matrix analysis is presented.  Section 4 is
devoted to summary and conclusions.  Throughout the paper, we use the
units in which the speed of light is unity.

\section{Theoretical Modeling}

In this analysis we consider a spatially--flat universe for simplicity,
consisting of baryons, cold dark matter, and dark energy. We ignore the
dark energy clustering, and assume that the spatial fluctuations
entirely originate from the matter component (i.e., baryons and dark
matter).  We further model that the cosmic expansion history {\it
effectively} follows the universe with the matter density parameter
$\Omega_{\rm m}$ and the dark energy parameter $1-\Omega_{\rm m}$:
\begin{eqnarray}
  H(a)^2=H_0^2\left[
  \Omega_m a^{-3}+(1-\Omega_m)a^{-3(1+w_0+w_a)}e^{3w_a(a-1)}
  \right],
\label{friedman}
\end{eqnarray}
where $H_0=100h~{\rm km~s^{-1}~Mpc^{-1}}$ is the Hubble constant, $a$ is
the cosmic scale factor, and $w_0$ and $w_a$ are constants parameterizing
the equation of state of dark energy\cite{CP,Linder2003,CMP}:
\begin{eqnarray}
\label{eq:eos}
 p/\rho \equiv  w(a)=w_0+w_a{(1-a)}.
\end{eqnarray}
Note that we use equation (\ref{friedman}) even in the DGP model that
does not have dark energy at all by approximating its cosmic expansion
law with the two parameters $w_0$ and $w_a$. In this case, they do not
have any relations to dark energy in reality, but it is already shown
that such an empirical description provides a reasonable approximation
to the cosmic expansion in the DGP model. For definiteness, the expansion
in the DGP has the {\it effective} equation of state (e.g., \cite{Linder2005})
\begin{eqnarray}
\label{eq:eosdgp}
 w(a)=-{1\over 1+\Omega_m(a)},
\end{eqnarray}
where
\begin{eqnarray}
\label{eq:omega_a}
  \Omega_m(a)={H_0^2\Omega_m a^{-3}\over H(a)^2}.
\end{eqnarray}
The cosmic expansion in the DGP model is well approximated
by the dark energy model with {\it effective} equation of state with 
$w_0=-0.78$ and $w_a=0.32$ as long as $\Omega_m\sim0.27$. 
The parameterization gives the distance redshift relation 
within $0.5$ \% out to the redshift $2$ \cite{Linder2005}.

\subsection{Linder's $\gamma$ parameter}

According to refs. \cite{Linder2005,HL,LC}, the linear growth factor 
in the DGP and dark energy models is well approximately expressed by
\begin{eqnarray}
\label{eq:lineargrowth}
 {D_1(a)\over a }\propto\exp\left[{\int_0^a {da'\over a'}\left(
\Omega_m(a')^\gamma-1\right)}\right].
\end{eqnarray}
In this description, the constant parameter $\gamma$ characterizes 
the gravity force model, i.e., the Poisson equation.  

The dark energy models with the effective
equation of state (\ref{eq:eos}) within the
general relativity are well approximated by 
\begin{eqnarray}
  && \gamma=0.55+0.05[1+w(z=1)] ~~~~~(w>-1),
\\
  && \gamma=0.55+0.02[1+w(z=1)] ~~~~~(w<-1).
\end{eqnarray}
This formula reproduce the exact linear growth factor within $0.3$\%
($0.5$\%) for $-1.2<w<-0.8$ $(-1.5<w<-0.5)$.  Therefore $\gamma$ in dark
energy models takes the value $\gamma=0.54-0.56$ for $-1.2<w<-0.8$
\cite{Linder2005,HL,LC}.

On the other hand, in the DGP model, the Poisson equation is modified in
the linear regime. Then $\gamma$ takes a different value from that of
the dark energy model even if the background expansion is same (i.e. if
$w_0$ and $w_a$ are same).  Ref. \cite{LC} found that in the DGP model 
$\gamma = 0.68$ is an excellent approximation for the evolution of the 
growth factor and that $\gamma$ varies by only $2$ \% into the past.

The point here is that a dark energy model mimicking the cosmic expansion
history of the DGP model predicts a different linear growth rate by
$\Delta \gamma \sim 0.1$. In what follows, therefore, we employ equations
(\ref{friedman}) to (\ref{eq:omega_a}) to describe the expansion history
and the growth of density fluctuations, which empirically describe both
the DGP and dark energy models, and ask if it is possible to achieve the
accuracy of $\Delta \gamma \sim 0.1$ by optimizing future surveys of
galaxies.

\subsection{Weak lensing power spectrum and Fisher matrix}

The optimization of imaging surveys is based on the weak lensing
tomography method (see e.g., \cite{TW,tomoHu,HTBJ,DJT}).  In this
methodology, one divides the entire galaxy samples in several different
redshift bins according to the weight factor $W_i(z(\chi))$ for the
$i$-th redshift bin:
\begin{eqnarray}
  W_i(z)={1\over \bar N_i} \int_{\max(z_i,z)}^{z_{i+1}} dz'
  {dN(z')\over dz'}\left(1-{\chi(z)\over \chi(z')}\right),
\end{eqnarray}
where $dN/dz$ denotes the differential number count of galaxies with
respect to redshift per unit solid angle (see below for details),
$\chi(z)$ is the radial comoving distance at $z$,
\begin{eqnarray}
  \chi(z)=\int_0^z{dz'\over H(z)}={1\over H_0}\int_0^z{dz'\over 
\sqrt{\Omega_m (1+z')^{3}+(1-\Omega_m)(1+z')^{3(1+w_0+w_a)}
e^{-3w_az'/(1+z')}}},
\label{chidefi}
\end{eqnarray}
and
\begin{eqnarray}
\bar N_i=\int_{z_i}^{z_{i+1}} dz'
  {dN(z')\over dz'}
\end{eqnarray}
is the total number of galaxies in the $i$-th redshift bin.  While
imaging surveys provide photometric redshifts alone from the multi-band
photometry, instead of spectroscopic redshifts, for galaxies, it is
known that the lensing tomography works even with relatively crude
redshift information.

Assuming that the anisotropic stress is negligible, the cosmic shear
power spectrum is given as:
\begin{eqnarray}
  P_{(ij)}(l)=\int d\chi W_i(z(\chi)) W_j(z(\chi)) 
  \left({3H_0^2 \Omega_m\over 2a}\right)^2
  P_{\rm mass}^{\rm Nonlinear}
  \left(k\rightarrow {l\over\chi},z(\chi)\right),
\end{eqnarray}
where $P_{\rm mass}^{\rm Nonlinear} \left(k,z\right)$ is the nonlinear
mass power spectrum at the redshift $z$, $k$ is the wave number of the
three dimensional coordinates, $l$ is the wave number of the two
dimension corresponding to the angular coordinates, $a$ is the scale
factor normalized to unit at the redshift $z=0$. We compute $P_{\rm
mass}^{\rm Nonlinear} \left(k,z\right)$ adopting the Peacock and Dodds
formula \cite{PD}.

The covariance matrix for $P_{(ij)}(l)$ is approximately given by
\begin{eqnarray}
 {\rm Cov}\bigl[P_{(ij)}(l),P_{(mn)}(l')\bigr]
&=&{\delta_{ll'}\over (2l+1)\Delta l f_{\rm sky}}
\bigl[P_{(im)}^{\rm obs}(l)P_{(jn)}^{\rm obs}(l)+
      P_{(in)}^{\rm obs}(l)P_{(jm)}^{\rm obs}(l)\bigr]
\nonumber
\\
&\equiv&\delta_{ll'}{\rm Cov}_{(ij)(mn)}(l),
\end{eqnarray}
where we define
\begin{eqnarray}
  {P^{\rm obs}_{(ij)}(l)}=P_{(ij)}(l)+\delta_{ij}
  {\sigma_\varepsilon^2\over \bar N_i},
\end{eqnarray}
$f_{\rm sky}$ is the fraction of the survey area, and
$\sigma_\varepsilon$ is the rms value of the intrinsic ellipticity of
randomly oriented galaxies, for which we adopt $\sigma_\varepsilon=0.4$
(see e.g., \cite{TW,tomoHu,HTBJ}).

Finally the Fisher matrix is estimated as
\begin{eqnarray}
F_{\alpha\beta}=\sum_l\sum_{(ij)(mn)} 
  {\partial P_{(ij)}(l)\over \partial\theta^\alpha}
  {\rm Cov}_{(ij)(mn)}^{-1}(l)
  {\partial P_{(mn)}(l)\over \partial\theta^\beta},
\end{eqnarray}
where $\theta^\alpha$ denote a set of parameters in the theoretical
modeling.  To be more specific, we consider 7 parameters, $\gamma$,
$w_0$, $w_a$, $\Omega_m$, $\sigma_8$ (the fluctuation amplitude at
$8h^{-1}$Mpc), $h$, and $n_s$ (the primordial spectral index of matter
power spectrum), assuming the other cosmological parameters are
determined from independent cosmological data analysis.

We adopt the range of $10\leq l\leq 10^4\times(N_g/35/n_b)^{1/2}$ for the
sum of $l$, where $N_g$ is the number density of galaxy per unit solid
angle (see next subsection).  We define the 3 dimensional Figure of
Merit by the reciprocal of the volume of the error ellipsoid enclosing
the 1 sigma confidence limit in the $\{\gamma,w_0, w_a\}$ space,
marginalizing the Fisher matrix over the other parameters.  Similarly,
the 2 dimensional Figure of Merit is the reciprocal of the surface of
the error ellipse enclosing the 1 sigma confidence limit in the $\{w_0,
w_a\}$ plane with $\gamma$ fixed.

\subsection{Modeling galaxy sample}


We assume the following form of the redshift distribution of the galaxy sample per
unit solid angle
\begin{eqnarray}
{dN\over dz}={N_g\beta\over z_0^{\alpha+1} \Gamma((\alpha+1)/\beta)}z^ 
\alpha
   \exp\left[-\left({z\over z_0}\right)^\beta\right],
\label{dndzdef}
\end{eqnarray}
where $\alpha$, $\beta$, and $z_0$ are the parameters, and
$N_g=\int dz dN/dz$.
The mean redshift may be determined by
\begin{eqnarray}
z_m= {1\over N_g}\int dz z{dN\over dz}
    ={z_0\Gamma((\alpha+2)/\beta)\over \Gamma((\alpha+1)/\beta)}.
\end{eqnarray}

We assume that $N_g$ and $z_m$ is related to the exposure
time $\texp$ as, following the reference \cite{OptimalLensingTom},
\begin{eqnarray}
&&z_m= 0.9\left({\texp\over 30 ~{\rm min.}}\right)^{0.067},
\label{scalezm}
\\
&&N_g=35 \left({\texp\over 30 ~{\rm min.}}\right)^{0.44} {\rm arcmin.} 
^{-2}
\label{scaleng}
\end{eqnarray}
The mean redshift $z_m$ changes from 0.72 to 1.1, and $N_g$ does from
$7.8$ to $163$, as the exposure time $\texp$ changes from $1$ minute to
$10^3$ minutes.  In the reference \cite{OptimalLensingTom}, $\alpha=2$
and $\beta=1.5$ are adopted. However, in the present paper, we adopt
$\alpha=0.5$ and $\beta=3$.

\begin{figure}
\begin{center}
    \leavevmode
    \epsfxsize=16cm
    \epsfbox[20 50 600 720]{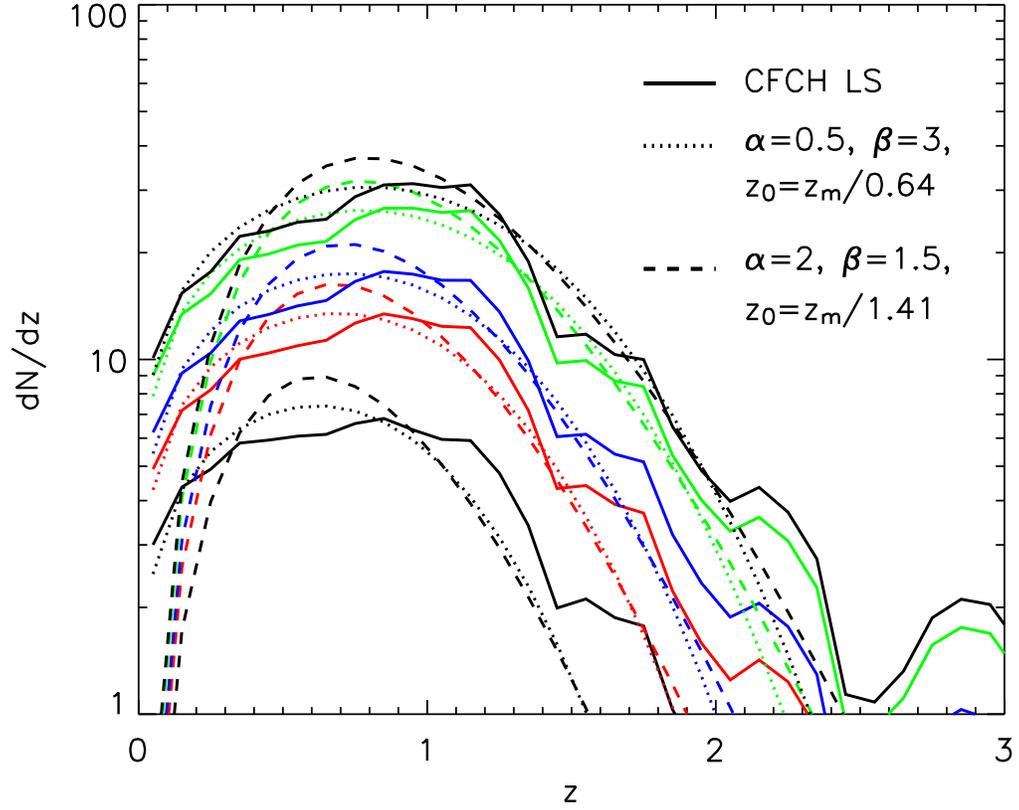}
\end{center}
\caption{$dN/dz$ as function of the exposure time, 
$\alpha=2,~\beta=1.5$ with $z_0=z_m/1.41$ (dashed curve), and
$\alpha=0.5,~\beta=3$ with $z_0=z_m/0.64$ (dotted curve),
respectively, for the fitting function of the form (\ref{dndzdef}),
for the exposure time $\texp=1,~5,~10,~30,~45$ minutes
from bottom to top. The solid curve shows the corresponding 
CFHT LS photo-z $i$ band data.
}
\label{figurea}
\end{figure}

In order to check the validity of our mock galaxy samples, we show in
Figure \ref{figurea} the two cases of $\alpha=0.5$ and $\beta=3$ (dotted curve), and
$\alpha=2$ and $\beta=1.5$ (dashed curve), for exposure times of $t_{\rm
exp}=1,~5,~10,~30,~45$ minutes (from bottom to top respectively).  The
solid curves show the real redshift histograms, for the corresponding
$i$band magnitude limits, taken from the CFHT photometric redshift data
of \cite{CFHT}. These photo-z's were calibrated using the VVDS
spectroscopy and are reliable to $i\simeq25$ which is sufficient for
this study (see \cite{CFHT}). The relationship between magnitude limit
and exposure time was scaled from the published Subaru Suprime-Cam data
of \cite{Miyazaki}. These data are shown in Table I for the $i,~g,~r,~z$
passbands.  Denoting the exposure time for the $i$ band by $\texp$, the
exposure time for $g$ band is about $\texp_g=3\times\texp$.  Similarly,
$\texp_r=1.2\times\texp$ for $r$ band, and $\texp_z=0.3\times\texp$ for
$z$ band, respectively.

\begin{table}
\begin{center}
\begin{tabular}{ccccc}
\hline
~~~~~$i_{\rm AB~limit}$~~~~~  & ~~~~~~ $i(S/N=10)$ ~~~~~~ & ~~~~~~ $g(S/N=5)$ ~~~~~~ & ~~~~~~ $r(S/N=5)$ ~~~~~~ &  
~~~~~~ $z(S/N=5)$ ~~~~~~  \\ 
\hline 
 $22.97$ &       $1$~mins. & $3$~mins. & $1.1$~mins. & $0.3$~mins.  \\
 $23.84$ &       $5$~mins. & $15$~mins. &  $7$~mins. & $1.4$~mins.   \\
 $24.22$ &       $10$~mins. & $30$~mins. & $12$~mins. & $3.5$~mins. \\
 $24.81$ &       $30$~mins. & $90$~mins. & $34$~mins. & $8.1$~mins.  \\
 $25.04$ &       $45$~mins. & $130$~mins. & $50$~mins. & $13$~mins.  \\
\hline
\end{tabular}
\end{center}
\caption{Exposure time for the bands, $i, g, r, z$.}
\end{table}

\begin{table}
\begin{center}
\begin{tabular}{c|c|c|c|c}
\hline
~~~~~${\rm Sub-sample}$~~~~~  & ~~~~~~~ $n_b=1$ ~~~~~~~ & ~~~~~~~ $n_b=2$ ~~~~~~~ & ~~~~~~~ $n_b=3$ ~~~~~~~ &  
~~~~~~~ $n_b=4$ ~~~~~~~~~~~  \\ 
\hline 
 choice of band  &       $i$ & $i,~r$ & $g,r,i,z$ & $g,r,i,z$  \\
\hline 
 $\sum_j\texp_j$ &       $\texp$ & $2.2\times \texp$ &  $5.5\times\texp$ &  $5.5\times\texp$ \\
\hline 
 redshift bins   &       $0.05<z<2.5$ & $0.05<z<z_m$  & $0.05<z<3z_m/4$  &  $0.05<z<0.6\times z_m$\\
        ~         &           ~         & $z_m<z<2.5$   & $3z_m/4<z<5z_m/4$ & $0.6\times z_m<z<z_m$\\
        ~         &           ~         &        ~      & $5z_m/4<z<2.5$   & $z_m<z<1.4\times z_m$\\
         ~        &           ~         &        ~      &             ~    & $1.4\times z_m<z<2.5$\\
\hline
\end{tabular}
\end{center}
\caption{Assumption on the subsample and measurement}
\end{table}

The total survey area can be expressed as
\begin{eqnarray}
  {\rm Area}=  \pi \left({{\rm Field~of~View}\over 2}\right)^2
       {T_{\rm total} \over 1.1\times \sum_{j} \texp_{\rm j}+t_{\rm op}},
\label{area}
\end{eqnarray}
where we assume that the Field of View of $1.5$ degree, the total
observation time $T_{\rm total}$ is fixed as $800$ hours, and the
overhead time is modeled by a constant, $t_{\rm op}=5$ minutes, plus a fraction  ($10\%$) of the exposure time $\sum_{j} \texp_{\rm j}$ for one field 
of view.

We consider the cases the tomography is used, which we denote by
$n_b=2$, $n_b=3$ and $n_b=4$. Here $n_b$ denotes the number of the
redshift bin.  In the case $n_b=2$, the sample is divided into the two
subsamples in the range $0.05<z<z_m$ and $z_m<z<2.5$, while in the case
$n_b=3$, we consider the three subsample $0.05<z<3z_m/4$,
$3z_m/4<z<5z_m/4$ and $5z_m/4<z<2.5$.  In the case $n_b=4$, we consider
the four subsample $0.05<z<0.6\times z_m$,~$0.6\times z_m<z<z_m$,
~$z_m<z<1.4\times z_m$, and $1.4\times z_m<z<2.5$ (see also Table II).
We also consider the case the tomography is not used, which we denote by
$n_b=1$, for which we don't take into account how to obtain $dN/dz$, instead
assuming that $dN/dz$ is obtained by some method.

We assume that the subsample of $n_b=2$ is constructed by the two band,
$r$ and $i$, observation, given that the strategy proposed in
\cite{CT} is successful.  The cases $n_b=3$ and $n_b=4$ are constructed
by the 4 band $g,~i,~r,~z$, observation, assuming that the conventional
photo-z is successful.  The case $n_b=1$ is based on the $i$ band
observation.  We assume that $90$\% galaxies of $i$ band measurements
$dN/dz$ can be used as the subsample, in the case $n_b=2,~3,~4$.

We use $\texp$ to represent the $i$ band exposure time for one field of
view, then we assume $\sum_{j} \texp_{\rm j}=5.5\times\texp$ for the
cases $n_b=3,4$, $\sum_{j} \texp_{\rm j}=2.2\times\texp$ for the case
$n_b=2$, and $\sum_{j} \texp_{\rm j}=\texp$ for the case $n_b=1$,
respectively.

\begin{figure}
\begin{center}
    \leavevmode
    \epsfxsize=16cm
   \epsfbox[20 50 600 720]{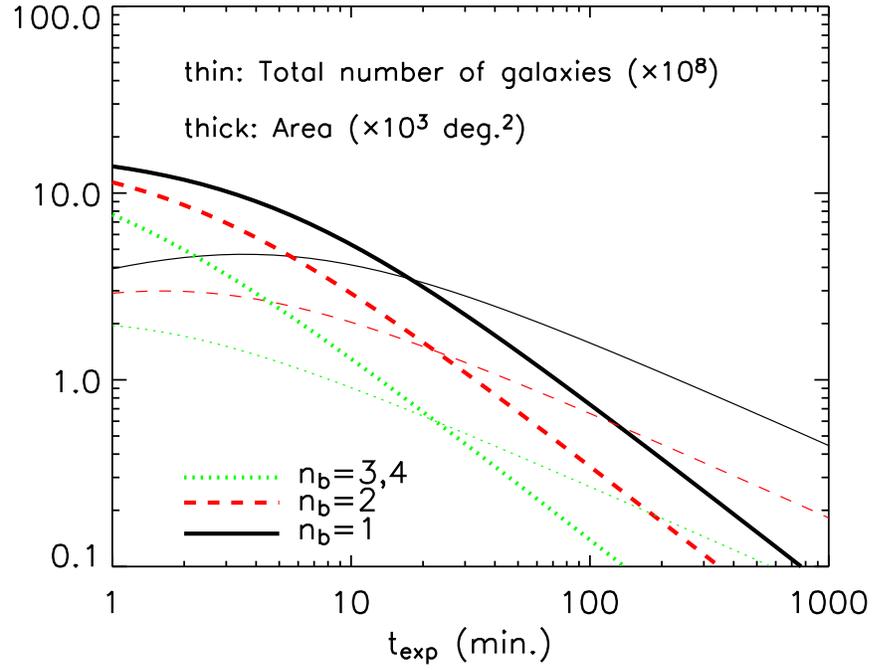}
\end{center}
\caption{The total survey area (thick), 
and the total number of the galaxies (thin) 
as function of the $i$ band exposure time $\texp$,
for the case $n_b=1,~2,~3~{\rm and}~4$.}
\label{figureb}
\end{figure}

Figure \ref{figureb} shows the resultant total survey area, and the total number of
galaxies as function of the $i$ band exposure time $\texp$, for the
cases, $n_b=1,~2,~3~{\rm and}~4$.

\subsection{DGP model}

Here we demonstrate the weak lens power spectrum of the dark energy
model and the DGP model with the same cosmic expansion.  The linear
perturbation theory in the DGP model has been extensively worked out by
\cite{KM}. While more recently Koyama and Silva studied nonlinear
evolution of density fluctuations in the DGP model \cite{KS}, the
nonlinear nature of the gravity in the DGP model is still an unsolved
problem.  Therefore we adopt an empirical modeling of the nonlinear
growth combining the Peacock-Dodds nonlinear fitting formula \cite{PD}
and the linear growth rate in the DGP model \cite{KM}. As a result, our
predictions below may be inaccurate on nonlinear scales, but our main
conclusions concerning the optimization strategy would be unlikely to be
sensitive to this approximation.

\begin{figure}
\begin{center}
    \leavevmode
    \epsfxsize=16cm
    \epsfbox[20 50 600 720]{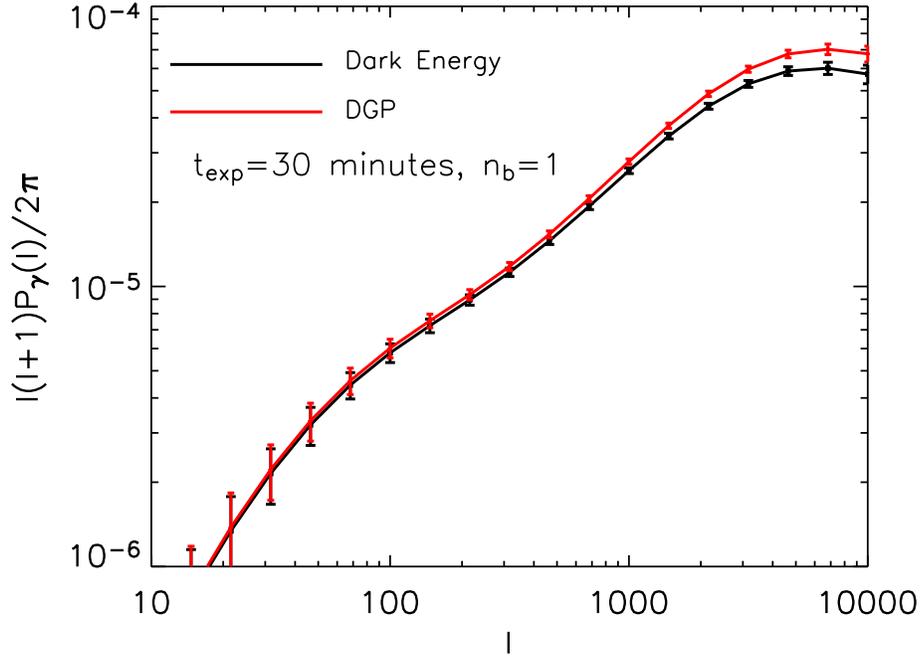}
\end{center}
\caption{
The dark (black) curve is the weak lensing power spectrum of 
the dark energy model with the cosmological parameter, 
$\Omega_m=0.27$, $\Omega_b=0.044$, $h=0.72$, 
$\sigma_8=0.8$, $n_s=0.95$, and the equation of state 
parameter of the dark energy $w_0=-0.78$, $w_a=0.32$, 
while the bright (red) curve is the flat DGP model of the 
same cosmological parameters. Here we assume the 
HSC like survey with $t_{\rm exp}=30$ minutes of the case 
$n_b=1$(see section 2 for details).}
\label{figurec}
\end{figure}

Figure \ref{figurec} shows the weak shear power spectrum of the spatially flat DGP
model and the dark energy model with the same background expansion.  The
cosmological parameters of both of the models are the same ($\Omega_m=0.27$,
$\Omega_b=0.044$, $h=0.72$, $\sigma_8=0.8$, and the spectral index
$n_s=0.95$).  To realize the same cosmic expansion history, the {\it
effective} equation of state parameter of the dark energy is chosen as
$w(z)=-0.78+0.32 z/(1+z)$, as mentioned in Section 2.1.  A similar
computation has been already considered by \cite{Ishak}, but our present
work differs in that we use the Peacock \& Dodds formula and
that we assume a rather shallow sample of galaxies.  Because the Poisson
equation of the DGP model is modified, then the difference comes from
the growth rate.  In this figure we assume $30$ minutes exposure time of
$n_b=1$.  The theoretical curves and the errors bar depend on the survey
sample, but we might expect that the two curves could be distinguished.
In the next section, we examine the capability of the differentiation.

\section{Results}

In this section, we present our optimization analyses for the HSC weak lensing survey. 
Specifically, we fix the total observation time of the HSC survey, $T_{\rm tot}$ 
as $800$ hours, and adopt a model 
of the background galaxy sample described in Sec. IIC for the HSC survey; 
in particular the mean redshift of galaxies $z_{\rm m}$ and their surface 
number density $N_{\rm g}$ are given by eqs.  (\ref{scalezm}) and
(\ref{scaleng}) as a function of the exposure time $\texp$.
In this section , we also present results in combination with 
a spectroscopic survey, the WFMOS BAO survey, 
which will be limited by a total observation time
(see \cite{David} for discussion of the optimization under 
this condition). 
Note that we also assume that the WFMOS survey is limited 
by the total survey area of the HSC imaging survey. 
Namely, the survey area of the WFMOS survey must be
less than or equal to that of the HSC survey, as the HSC survey is 
acting as a photometric source catalogue for the WFMOS spectroscopic survey. 
So, for the WFMOS survey, we fix the same survey area as the HSC 
imaging survey equation (\ref{area}), and the redshift range 
of galaxies $0.8\leq z\leq 1.4$ with the number density $\bar
n=4\times10^{-4}$ ${h^3}$Mpc${}^{-3}$ \cite{David}, which
is a set of optimized survey parameters for the spectroscopic survey. 

\begin{figure}
\begin{center}
    \leavevmode
    \epsfxsize=16cm
    \epsfbox[20 50 600 720]{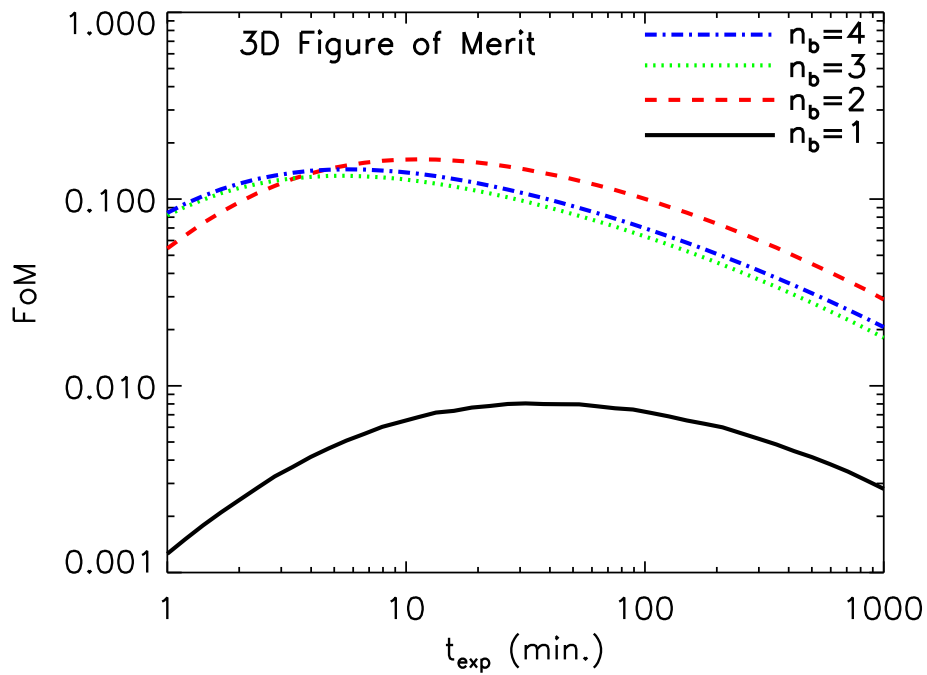}
\end{center}
\caption{
Three dimensional (3D) FoM in $\{\gamma,w_0, w_a\}$ as function of the
$i$ band exposure time, which is obtained from the Fisher matrix of the 7 
parameters $\gamma, w_0, w_a, \Omega_m, \sigma_8, h$, and $n_s$, 
Here the target parameter is 
$\gamma=0.55$, $w_0=-1$, $w_a=0$, $\Omega_m=0.27$, $\sigma_8=0.8$, 
$h=0.72$, $n_s=0.95$. The other parameter, $\Omega_b=0.044$ is fixed. }
\label{figured}
\end{figure}

Figure \ref{figured} shows the Figure of Merit (FoM) of the 3 dimension (3D) of
$\{\gamma,w_0, w_a\}$, as function of the exposure time, $\texp$.  
The 3D FoM, the reciprocal of the volume of the 1$\sigma$ error 
ellipsoid in the $\{\gamma,w_0, w_a\}$ space, is computed by 
marginalizing the Fisher matrix of the 7 parameters over 
$\Omega_m, \sigma_8, h$, and $n_s$, with a fixed value for the 
baryon density, $\Omega_b=0.044$.

The lensing tomography method with $n_b=2$ significantly improves the 3D
FoM, and continues to do so with increasing $n_b$ for $\texp \simlt
10$mins.  The peak of the FoM systematically shifts to the shorter
exposure time with larger $n_b$, while the peak profile is fairly broad.
With increasing $n_b$, more information of redshift evolution of structure
can be obtained. Similarly, as $\texp$ increases, more information of
smaller structure can be obtained. However, these are offset by decrease
in total survey area.  Namely, observation of more bands and longer
exposures consume observation time, and the total survey area
becomes smaller.  This decreases the FoM.

\begin{figure}
\begin{center}
    \leavevmode
    \epsfxsize=16cm
    \epsfbox[20 50 600 720]{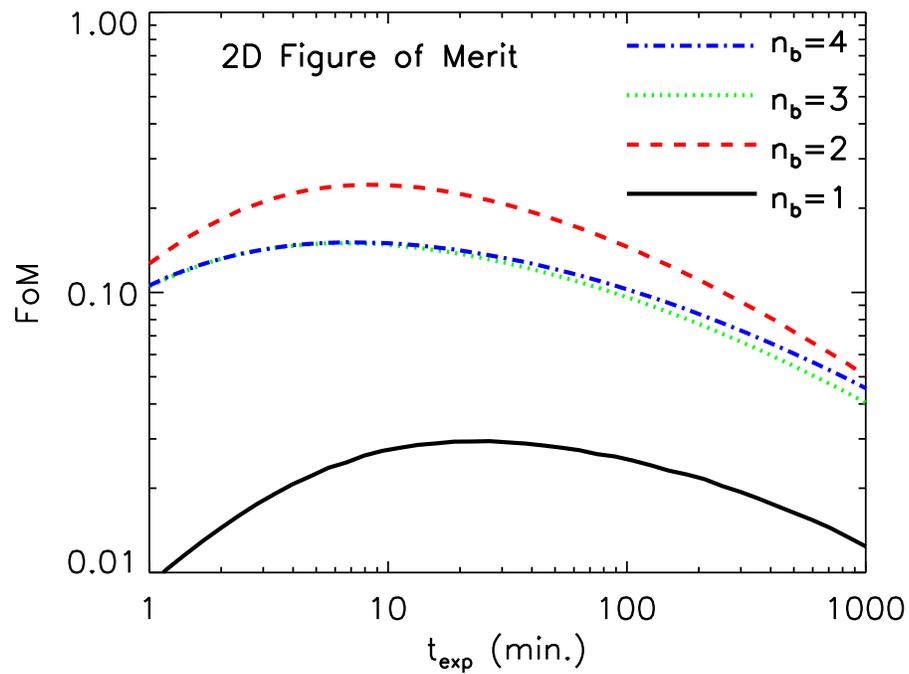}
\end{center}
\caption{
Two dimensional FoM in $\{w_0, w_a\}$ from the Fisher matrix of the 
6 parameters $w_0, w_a, \Omega_m, \sigma_8, h$. 
Here the fiducial model is $\Lambda$CDM, with  
$w_0=-1$, $w_a=0$, $\Omega_m=0.27$, $\sigma_8=0.8$, 
$h=0.72$, $n_s=0.95$. The other parameters, $\gamma=0.55$ and $\Omega_b=0.044$ are fixed. }
\label{figuree}
\end{figure}

For comparison, we plot in Figure \ref{figuree} the 2D FoM, the reciprocal of the
area of the 1$\sigma$ error ellipse in the $\{w_0, w_a\}$ plane,
evaluated by marginalizing the Fisher matrix of the 6 parameters ($w_0,
w_a, \Omega_m, \sigma_8, h$, and $n_s$) with $\Omega_b=0.044$ and
$\gamma=0.55$ fixed. One can find the similar features as those of the 3D
FoM.  This figure suggests the three redshift bin is enough to constrain
$w_0$ and $w_a$ and that the peak of FoM is located around $\texp
\approx$ $10$ minutes, and the peak profile is very broad.  The FoM of
the case $n_b=2$ is larger than that of $n_b=3,4$.  This indicates that
observation of larger survey area with small number of bands ($n_b=2$)
can be useful for the dark energy constraints, though an accurate-photo-z
strategy is required.

\begin{figure}
\begin{center}
    \leavevmode
    \epsfxsize=16cm
     \epsfbox[20 50 600 720]{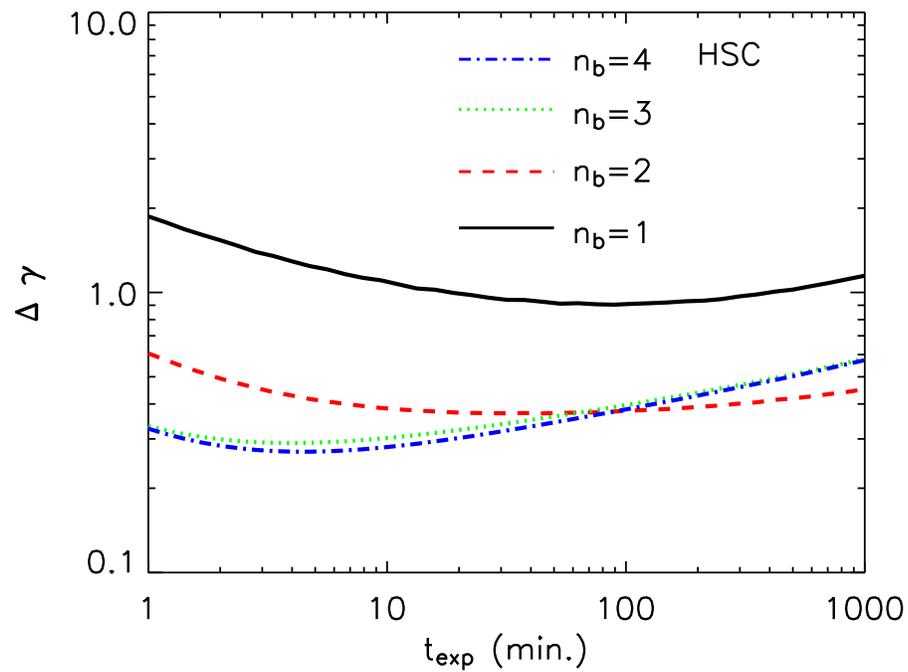}
\end{center}
\caption{
$1$ sigma error in measuring $\gamma$ as function of 
the exposure time,
obtained by marginalizing the Fisher matrix of the $7$ parameters
$\gamma, w_0, w_a, \Omega_m, \sigma_8, h$, and $n_s$, over the 
parameters other than $\gamma$. The target parameters 
is the same as those of Figure \ref{figured}.
}\label{figuref}
\end{figure}

Figure \ref{figuref} shows the 1$\sigma$ error on $\gamma$ as a function of $\texp$,
which is estimated by marginalizing the Fisher matrix of the 7
parameters, $\gamma,w_0, w_a, \Omega_m, \sigma_8, h$ and $n_s$, over the
parameters other than $\gamma$. The curve shows the error from the weak
lensing power spectrum adopting a proposed survey with HSC; $\Delta
\gamma \approx 0.3 (1)$ can be achieved with (without) tomography.  The
result indicates that the weak lensing survey alone cannot reach the
accuracy of $\Delta \gamma =0.1$ that is required to distinguish between
the DGP and dark energy models.

\begin{figure}
\begin{center}
    \leavevmode
    \epsfxsize=16cm
    \epsfbox[20 50 600 720]{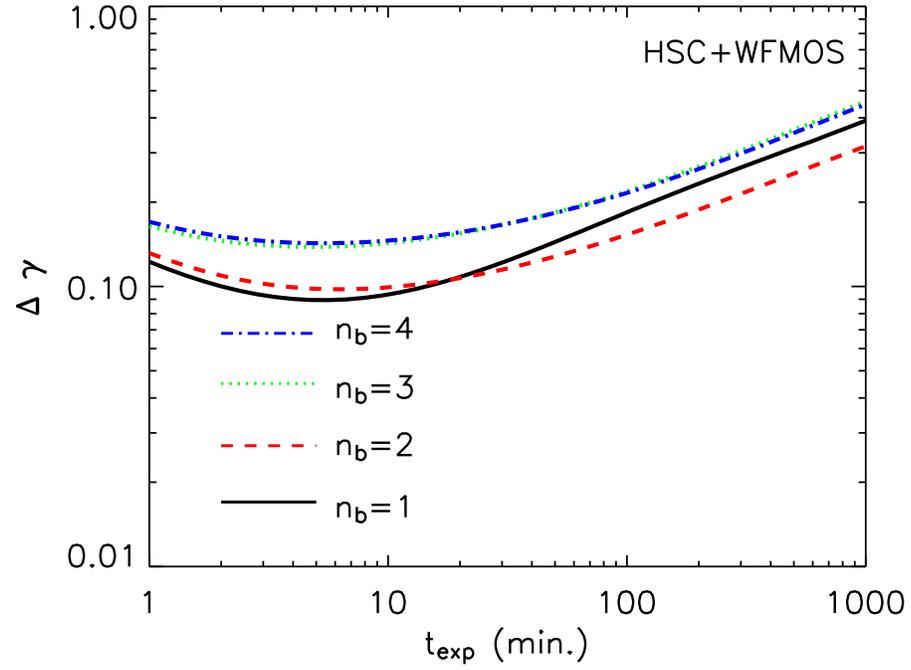}
\end{center}
\caption{
Same as figure \ref{figuref}, but the considering the case of the weak lensing 
power spectrum combined with the galaxy power spectrum of the 
redshift survey. 
}
\label{figureg}
\end{figure}

The uncertainty in $\gamma$ can be significantly (more than a factor of
three) reduced by combining the baryon oscillation features from the
WFMOS survey (Figure \ref{figureg}).  In modeling the galaxy power spectrum of the
redshift survey, we simply considered the linear theory specified by the
9 parameters $\gamma, w_0, w_a, \Omega_m, \sigma_8, h, n_s, b_0$ and
$p_0$, where $b_0$ and $p_0$ are the parameters for the bias model, for
which we adopted the scale independent bias model with the form
\begin{eqnarray}
b(z)=1+(b_0-1)(1+z)^{p_0}.
\label{biasmodel}
\end{eqnarray} 
Here we assumed the target parameters $b_0=1.38$ and $p_0=1$.  For the
theoretical modeling of the galaxy power spectrum and the computation of
the Fisher matrix, the range of the wavenumber $0.01~h$Mpc${}^{-1}$$\leq
k\leq 0.2~h$Mpc${}^{-1}$ is included, (see Appendix for details).

{}From Figures \ref{figuref} and \ref{figureg}, the error of $\gamma$ has a minimum of $\texp$
between several minutes and $100$ minutes, depending on the strategy.
For the weak lensing survey (HSC) alone, the tomography technique is
very effective in reducing the error, and the result is fairly
insensitive to the the choice of $\texp$.  An additional spectroscopic
survey (WFMOS) significantly reduces the error.  In this case, shallow
surveys with $\texp <10$ minutes provide the minimum error for $\gamma$.
Especially, the case $n_b=1$ and $n_b=2$ is significantly improved by
the combination.  This behaviour is understood as follows.  We assume
the total observation time of the WFMOS survey is not fixed, while
adopting the same survey area as the HSC survey.  Then, in these
figures, the cases $n_b=1$ and $n_b=2$ assumes larger survey area for
the redshift survey than that of the cases $n_b=3$ and $n_b=4$.
However, note that the minimum is located around the several minutes of
the exposure time even for the case $n_b=3$ and $4$.  Therefore, when
considering the combination with the redshift survey,
wider and shallower surveys are indeed prefered.

\begin{figure}
\begin{center}
    \leavevmode
    \epsfxsize=16cm
    \epsfbox[20 50 600 720]{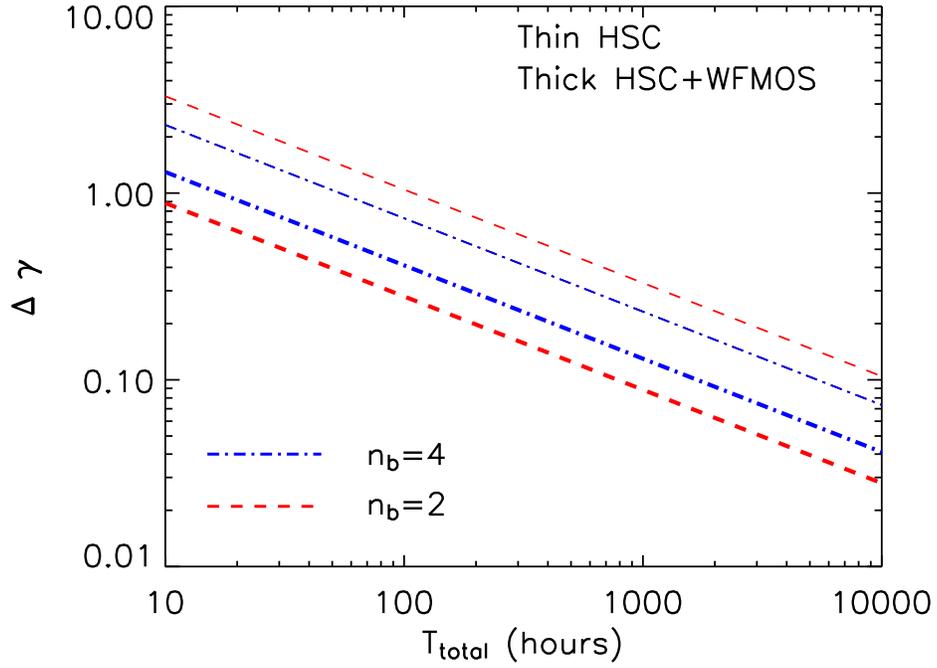}
\end{center}
\caption{
1 sigma error on $\gamma$ as function of the total observation time.
Here we fixed $\texp=10$ minutes and $n_b=4$(dash-dotted curve) and $n_b=2$ 
(dashed curve). The thin curve is the result with the 7 parameters 
of the Fisher matrix for the lensing power spectrum, but
the thick curve is the constraint from the combined weak lensing 
power spectrum and galaxy power spectrum (from a redshift survey).
}
\label{figureh}
\end{figure}

Now we are in a position to answer the question: is it possible to
distinguish between the DGP and dark energy models ?  For that purpose,
$\Delta \gamma\simlt 0.1$ is required. Figure \ref{figureh} plots the 1 sigma error
as a function of the total observation time $T_{\rm total}$, where we
adopt $t_{\rm exp}=10$ minutes and $n_b=4$ (dash-dotted curve) and
$n_b=2$ (dashed curve).  The thin curve is the result of the weak
lensing survey alone, while the thick curve is the result combined with
the redshift survey.  Note that $\Delta \gamma$ is in proportion to
$T_{\rm total}^{-1/2}$.  Figure \ref{figureh} suggests that the HSC survey alone
may reach $\Delta \gamma < 0.1$ with $T_{\rm tot}=10^4$hours,
the combination with the WFMOS survey may do so with
$T_{\rm tot}=10^3$hours if we put a prior constraint on $\Omega_b$.

\begin{figure}
  \leavevmode
  \begin{center}
    \begin{tabular}{ c c }
      \includegraphics[width=2.5in,angle=0]{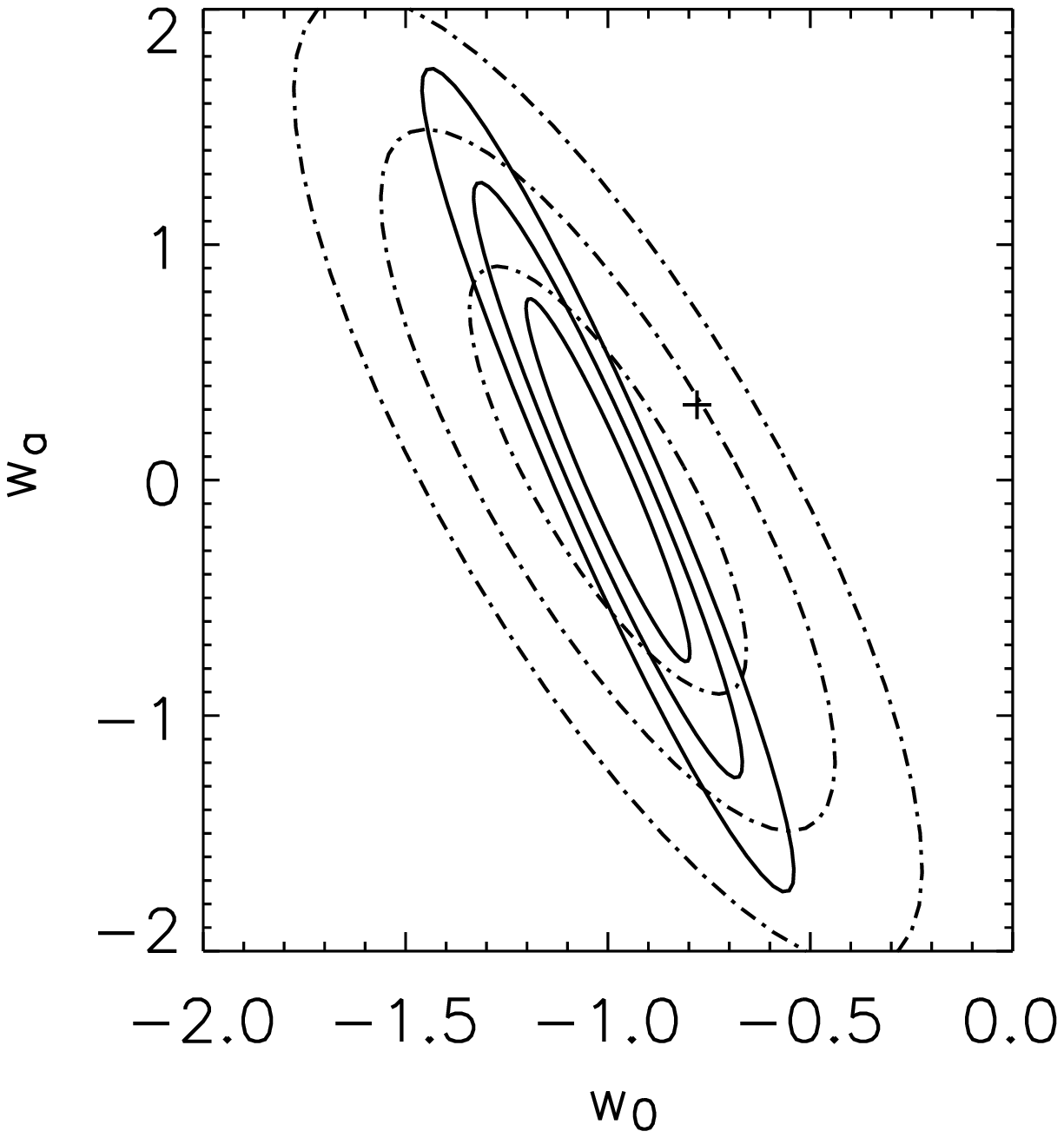}
      &
      \includegraphics[width=2.5in,angle=0]{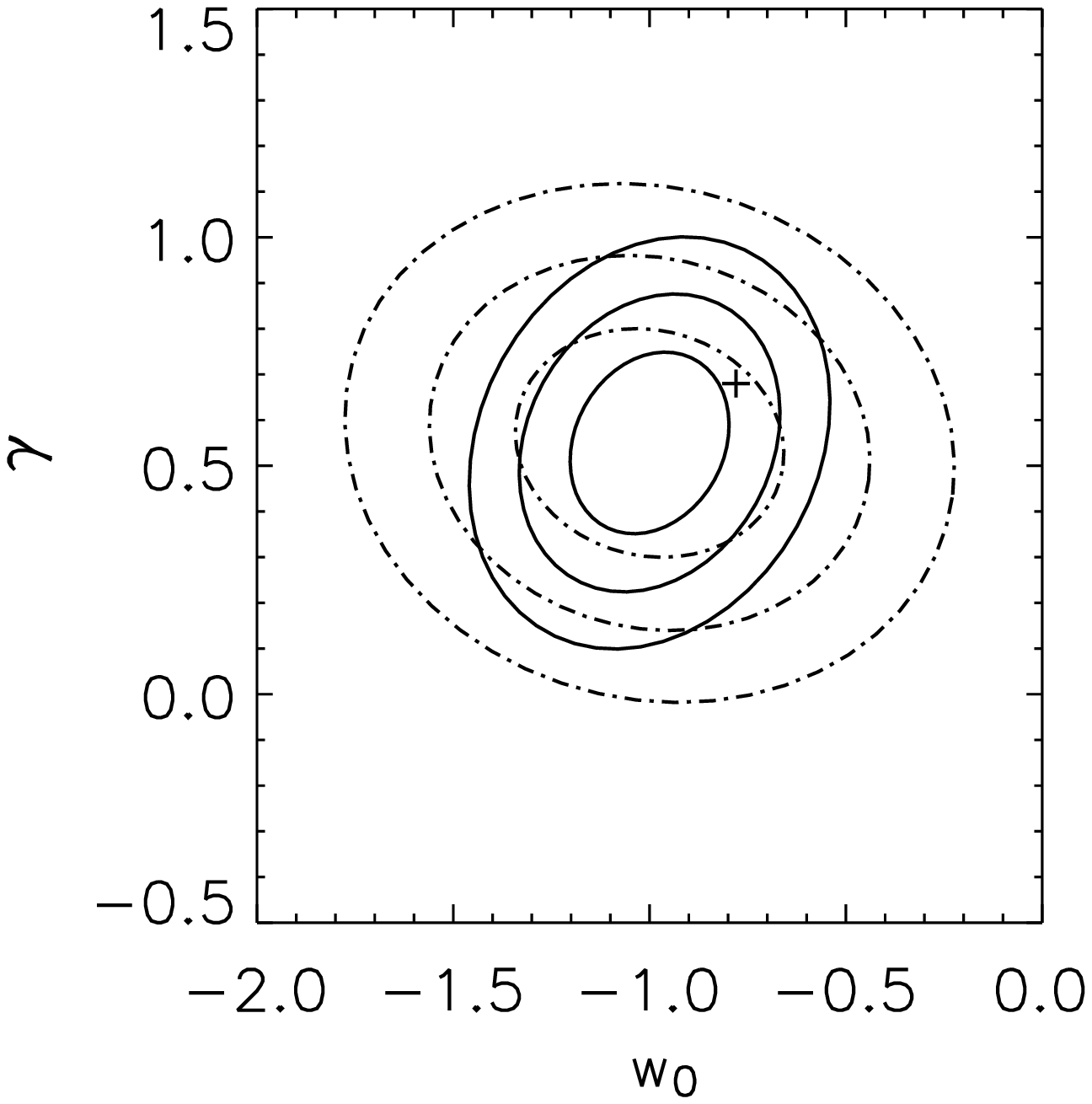}
    \end{tabular}
\caption{ (a,~Left) The $1$, $2$ and $3$-sigma contours 
in the $w_0-w_a$ plane. The dash-dotted curve is the result 
with the 7 parameters of the Fisher matrix for the lensing 
power spectrum and the Planck prior constraint, and
the solid curve is these constraints combined with the galaxy power 
spectrum from a redshift survey.
The target model is the $\Lambda$ CDM model, then $(w_0,w_a)=(-1,0)$, 
and the mark $(w_0,w_a)=(-0.78,0.32)$ is the DGP model.
Here we fixed $n_b=4$, $\texp=10$ minutes, and the total observation time, 
$800$ hours.
(b,~Right) Same as (a), but with the contours in the $w_0-\gamma$ plane from 
marginalizing the Fisher matrix of the 7 parameters over all other
parameters. 
The target model is the $\Lambda$CDM model, then $(w_0,\gamma)=(-1,0.55)$, 
and the mark $(w_0,\gamma)=(-0.78,0.68)$ is the DGP model.}
\label{figurei}
\end{center}
\end{figure}

Finally in this section, let us consider other impact that the HSC
survey may present as a test of modified gravity models. The dash-dotted
curves in Figure \ref{figurei}(a) show the $1$, $2$ and $3$-sigma confidence contours (going from the innermost outward) in the $w_0-w_a$ plane, by 
marginalizing the
Fisher matrix of the 7 parameters, $\gamma,w_0, w_a, \Omega_m, \sigma_8,
h$ and $n_s$, over the parameters other than $w_0$ and $w_a$.  
Here the constraint from future Planck survey is taken into account by
including the prior constraints $\Delta
\Omega_m=0.035,~\Delta\sigma_8=0.04,~ \Delta w_0=0.32,~\Delta
w_a=1,~\Delta n_s=0.0035$ \cite{Wang}.  
Here the target parameters are
same as those of the $\Lambda$CDM model in Figure \ref{figured}, and we fixed
$n_b=4$ and $\texp=10$ minutes.  Note that the point of the DGP model
$(w_0,w_a)=(-0.78,0.32)$ is marked, and is almost near the $2$ sigma
curve. This means that the HSC can distinguish between the DGP model and
the $\Lambda$CDM model at the $2$ sigma level by including future
constraint by the observation of the cosmic microwave background
anisotropy. Here, we fixed the total observation time as $800$ hours,
then the constraint can be improved when the total observation time is
longer.  The solid curve is the combination with the WFMOS survey, which
also shows the significant improvement of the constraint.
Similarly, figure \ref{figurei}(b) show the$1$, $2$ and $3$-sigma confidence 
contours in the $w_0-\gamma$ plane, by marginalizing the
Fisher matrix over the parameters other than $w_0$ and $\gamma$. 
The point of the DGP model $(w_0,\gamma)=(-0.78,0.68)$ is marked.
With this figure, the constraint is at the 1 sigma level. Then 
we can not clearly distinguish between the DGP model and the 
$\Lambda$CDM model with this plot. 
These features reflect how the shear power spectrum is sensitive to 
the parameters. This suggests the choice of a projection is 
important for distinguishing between these models.

\section{Summary and Conclusions}

In this paper, we investigated optimization of a weak lensing survey for
the dark energy, and how such a survey might be used for testing
modification of the theory of gravity. By introducing a simple model of
the survey sample as a function of the exposure time for one band of one
field of view, we investigated how the FoM and the constraint on
Linder's $\gamma$ parameter depend on the exposure time and the number
of passbands.  To optimize the survey to probe probe modifications of
gravity, we considered a Figure of Merit in the space
$\{\gamma,w_0,w_a\}$ as well as in the familiar 2D plane
$\{w_0,w_a\}$. We obtained the following results: 1)~The peak of the FoM
is located at $\texp\simeq {\rm several}\sim 10$ minutes for
$n_b=2,3,4$, though the peak profile is very broad.  2)~The tomography
technique improves the FoM effectively when including the parameter
$\gamma$.  3)~The combination with the redshift survey like the WFMOS
BAO survey improves the error on the parameter $\gamma$.  4)~The shallow
and wide survey is advantageous for the tomography, and has potential
when taking combination with the redshift survey into account. 5)~The
HSC weak lensing survey by itself is not sufficient for distinguishing
between the DGP model and a dark energy model with the same background
expansion, but it will be able to distinguish between the DGP and
$\Lambda$CDM at the 2 sigma level by including the prior constraint from
future CMB observation.

We assumed a very simplified model of the survey galaxy sample, and 
the error in the photometric redshift measurement is not taken into
account. Also we assumed that the weak lensing power spectrum of the
$10\leq l\leq10^4 (N_g/35/n_b)^{1/2}$ can be used.  Further
investigation is needed including the modeling of the galaxy
sample and the error in measuring the photometric redshift. 
In the present paper, we assumed the spatially flat universe.
In general, since the lensing power spectrum is not very sensitive to
the curvature of the universe, then the inclusion of the other
parameter will degrade the constraint \cite{B}.

\vspace{2mm}
\begin{acknowledgments}
This work is supported in part by Grant-in-Aid for Scientific research
of Japanese Ministry of Education, Culture, Sports, Science and
Technology (Nos.~18540277, 18654047, 18072002, 17740116, and 19035007), and by JSPS (Japan
Society for Promotion of Science) Core-to-Core Program ``International
Research Network for Dark Energy''. We thank M. Takada, S. Miyazaki,
H. Furusawa, K. Koyama, B. M. Schaefer, R. Maartens, B. A. Bassett, and
M. Meneghetti for useful comments related to the topic in the present
paper. We are also grateful to A. Taruya, T. Nishimichi, H. Ohmuro, 
K. Yahata, A. Shirata, S. Saito, M. Nakamichi and H. Nomura 
for useful discussions related to the topic in the present paper.
K.Y. is grateful to the people at Institute of Cosmology and
Gravitation of Portsmouth University for their hospitality and useful
discussions during his stay.
\end{acknowledgments}

\newpage
\begin{appendix}

\section{Modeling of the redshift survey power spectrum}

Here we briefly review the power spectrum and the Fisher matrix formula
for a galaxy redshift survey \cite{p2m,YBN}, adopted in the present
paper.  Here we assume a measurement of the multipole power spectrum
${\cal P}_l(k)$ $~(l=0,2)$ from the galaxy redshift survey, which we
theoretically model as
\begin{eqnarray}
  {\cal P}_{l}(k)=
  {1\over 2}\int d{{\mu}}
  {\int d{{\bf s}} \bar n({{\bf s}})^2{{\psi}}({{\bf s}},{{k}},\mu)^2 
  P({{k,\mu}},{{s}})
  {\cal L}_{l}(\mu)
\over
  \int d{{\bf s}}' \bar n^2({{\bf s}}') {{\psi}}({{\bf s}}',{{k}},\mu)^2},
\label{expll}
\end{eqnarray}
where $\bf s$ is the coordinate of the redshift space, $\bar n({\bf s})$
is the mean number density per unit volume, $\psi({{\bf s}},{{k}},\mu)$
is the weight factor, ${\cal L}_{l}({{\mu}})$ is the Lenegdre
polynomial, $\mu$ is the directional cosine between ${\bf k}$ and ${{\bf
s}}$, and $P({{k}},\mu,{{s[z]}})$ is the power spectrum at the redshift
$z$, which is modeled as
\begin{eqnarray}
  P(k,\mu,s[z])={s(z)^2\over \chi(z)^2}{ds(z)\over d\chi(z)}P_{gal}\left(
{q_{\scriptscriptstyle \|}}\rightarrow k\mu{ds(z)\over d\chi(z)},
{q_{\scriptscriptstyle \bot}}\rightarrow k\sqrt{1-\mu^2}{s(z)\over \chi(z)},z
\right)
\nonumber
\\
\end{eqnarray}
with 
\begin{eqnarray}
P_{gal.}({q_{\scriptscriptstyle \|}},{q_{\scriptscriptstyle \bot}},z)
  = b(z)^2\left[1+{d\ln D_1(z)/d\ln a(z)\over b(z)}{{q_{\scriptscriptstyle \|}}^2
  \over q^2}\right]^2 P_{\rm mass}^{\rm Linear}(q,z) 
\label{lin}
\end{eqnarray}
where $q^2={q_{\scriptscriptstyle \|}}^2+{q_{\scriptscriptstyle
\bot}}^2$, $P_{\rm mass}^{\rm Linear}(q,z)$ is the linear mass power
spectrum at the redshift $z$.  The comoving distance $\chi[z]$ is given by
\begin{eqnarray}
  \chi(z,\Omega_m,w_0,w_a)
={1\over H_0}\int_0^z{dz'\over 
\sqrt{\Omega_m (1+z')^{3}+(1-\Omega_m)(1+z')^{3(1+w_0+w_a)}
e^{-3w_az'/(1+z')}}},
\label{defsz}
\end{eqnarray}
as given in equation (\ref{chidefi}).  For our fiducial model we
adopt the flat $\Lambda$CDM model with $\Omega_m=0.27$. Thus, our
fiducial model is $s(z)=\chi(z,0.27,-1,0)$.  In the modeling of the bias,
we consider the scale independent bias model in the form,
Eq.(\ref{biasmodel}).

The variance of ${\cal P}_{l}(k)$ is given by 
\begin{eqnarray}
   \Delta {\cal P}_{l}(k)^2 &=&
  2{(2\pi)^3\over \Delta V_{k}}
  {\cal Q}^2_{l}({{\bf s}},k),
\label{defPPF}
\end{eqnarray}
where $\Delta V_{k}$ denotes the volume of the shell in the Fourier
space, and we have defined
\begin{eqnarray}
  &&{\cal Q}^2_{l}(k) = {1\over 2}
  \int d\mu   {
  \int d{{\bf s}} \bar n({{\bf s}})^4 
 {{\psi}}({{\bf s}},{{k}},\mu)^4 
  \bigl[P\bigl({{{k}}},\mu,{{s}}\bigr)
  +{1/\bar n({{\bf s}})}\bigr]^2
  [{\cal L}_{l}(\mu)]^2
  \over
  [\int d{{\bf s}}' \bar n({{\bf s}}')^2 {{\psi}}({{\bf s}}',{{k}},\mu)^2]^2}.
\label{defPPG}
\end{eqnarray}
Then, we may evaluate the fisher matrix by 
\begin{eqnarray}
  F_{\alpha\beta}\simeq\sum_{l=0,2}
  {1\over 4\pi^2} \int_{k_{\rm min}}^{k_{\rm max}} 
  \left[{\cal Q}^2_{l}(k) \right]^{-1}
  {\partial {\cal P}_{l}(k)\over \partial \theta^\alpha}
  {\partial {\cal P}_{l}(k)\over \partial \theta^\beta}
  k^2 dk.
\end{eqnarray}
In the present paper, we adopt $\bar n(s[z])=4\times10^{-4}$
${h^3}$Mpc${}^{-3}$ and $\psi({{\bf s}},{{k}},\mu)=1$.

\end{appendix}

\newpage
\def\and{{and }}




\begin{thebibliography}{99}
\bibitem{DGP}
  G. Dvali, G. Gabadadze, \and M. Porrati, Phys. Lett. B {\bf 485}, 208 (2000)
\bibitem{GG}
  G. Gabadadze, Astrophys. J. {\bf 597}, 566 (2003) 
\bibitem{YBRSY}
  K. Yamamoto, B. A. Bassett, R. C. Nichol, Y. Suto, \and K. Yahata,
  Phys. Rev. D {\bf 74}, 063525 (2006) 
\bibitem{Heavens}
  A. F. Heavens, T. D. Kitching, \and L. Verde, arXiv:astro-ph/0703191
\bibitem{Kunz} 
  M. Kunz\and D. Sapone,  arXiv:astro-ph/0612452 
\bibitem{Luty} 
  M. A. Luty, M. Porrati, \and R. Rattazzi, JHEP {\bf  0309}, 029 (2003) 
\bibitem{Nicolis} 
  A. Nicolis, \and R. Rattazzi, JHEP {\bf 0406}, 059 (2004)
\bibitem{ghostkoyama} 
  K. Koyama, Phys. Rev. D {\bf 72}, 123511 (2005) 
\bibitem{MM}
  R. Maartens \and E. Majerotto, Phys. Rev. D {\bf 74}, 023004 (2006)
\bibitem{SSH} 
  Y-S. Song, I. Sawicki, \and W. Hu, arXiv:astro-ph/0606286
\bibitem{KM}
  K. Koyama \and R. Maartens, JCAP {\bf 01}, 016 (2006) 
\bibitem{DETF}
  A. Albrecht et al., arXiv:astro-ph/0609591 
\bibitem{RESA}
  J. A. Peacock et al., arXiv:astro-ph/0610906 
\bibitem{OptimalLensingTom}
  A. Amara  \and A. Refegier, arXiv:astro-ph/0610127
\bibitem{David}
  D. Parkinson, C. Blake, M. Kunz, B. A. Bassett, R. C. Nichol, \and  K. Glazebrook,
  arXiv:astro-ph/0702040 
\bibitem{CP}
  M. Chevallier \and D. Polarski, Int. J. Mod. Phys. D {\bf 10}, 213 (2001)
\bibitem{Linder2003}
  E. V. Linder, Phys. Rev. Lett. {\bf 90}, 091301 (2003)
\bibitem{CMP}
  R. Crittenden, E. Majerotto, \and F. Piazza, arXiv:astro-ph/0702003
\bibitem{Linder2005}
  E. V. Linder, Phys. Rev. D {\bf 72}, 043529 (2005)
\bibitem{HL}
  D. Huterer \and E. V. Linder, arXiv:astro-ph/060868 
\bibitem{LC}
  E. V. Linder \and R. N. Cahn, arXiv:astro-ph/0701317
\bibitem{TW}
  M. Takada \and M. White, Astrophys.J. {\bf 601},  L1 (2004)
\bibitem{tomoHu} 
  W. Hu, Astrophys. J. 522, L21 (1999)
\bibitem{HTBJ}
  D. Huterer, M. Takada, G. Bernstein, \and B. Jain, MNRAS {\bf 366}, 101 (2006)
\bibitem{DJT}
  D. Dolney, B. Jain, \and M. Takada, MNRAS {\bf 366}, 884 (2006)
\bibitem{PD}
  J. A. Peacock \and S. J. Dodds, MNRAS {\bf 280}, L19 (1996)
\bibitem{CFHT}
  O. Ilbert, et al., A\&A {\bf 457}, 841 (2006)
\bibitem{Miyazaki}
  S. Miyazaki, et al., Publ.~Astron.~Soc.~Jap. {\bf 54}, 833 (2002)
\bibitem{CT}
  B. Jain, A. Connolly,  \and M. Takada, arXiv:astro-ph/0609338
\bibitem{KS}
  K. Koyama \and  F. P. Silva, arXiv:hep-th/0702169 
\bibitem{Ishak}
  M. Ishak, A. Upadhye \and D. N. Spergel, Phys. Rev. D {\bf 74} 043513 (2006)
\bibitem{Wang}
  S. Wang, J. Khoury, Z. Haiman, \and M. May, Phys. Rev. D {\bf 70}, l23008 (2004)
\bibitem{B}
  C. Clarkson, M. Cortes, \and B. A. Bassett, arXiv:astro-ph/0702670
\bibitem{p2m}
  K. Yamamoto, M. Nakamichi, A. Kamino, B. A. Bassett, \and H. Nishioka,
  Publ.~Astron.~Soc.~Jap. {\bf 58}, 93 (2006)
\bibitem{YBN}
  K. Yamamoto, B. A. Bassett, \and H. Nishioka, 
  Phys.~Rev.~Lett. {\bf 94}, 051301 (2005)
\end{thebibliography}
\end{document}